# SEARCH OF PARITY VIOLATION EFFECTS IN NEUTRON REACTION ON NATURAL LEAD

A. I. OPREA, C. OPREA, P.V. SEDYSHEV, YU. M. GLEDENOV

*Frank Laboratory for Neutron Physics, Joint Institute for Nuclear Research (FLNP-JINR), 141980 Dubna, Russian Federation*

*Abstract.* Parity violation effects (PV) in nuclear reaction were discovered in the '60 years of the last century in the capture of thermal transversal polarized neutrons by $^{113}$Cd nucleus. In this reaction experimentally was measured a non zero asymmetry of emitted gamma quanta and the results was interpreted by the existence of weak non leptonic interaction between nucleons in the compound nucleus. This first experimental result gave a serious impulse of theoretical and experimental developments of parity violation question in nuclear reactions.
  The weak interaction acts in the background of strong interaction (with order of magnitude higher) and therefore it is very difficult to observe and evidence it. One possibility is the evaluation of asymmetry effects induced by PV phenomena.
  For neutrons scattering there are a few asymmetry effects (like polarization of incident neutron beam, spin rotation and emitted neutrons asymmetry of incident transversal polarized neutrons) explained by the presence of weak interaction.
  In natural Lead were observed an unexpected high value of neutron spin rotation due to the PV phenomena. The natural Lead contains four isotopes and the main contribution to the PV effects is given by $^{204}$Pb. Further to explain the high value of neutron spin rotation it was supposed the existence of a new negative P resonance with energy $E_P = -16$ eV.
  In this work were estimated the PV effects in neutrons scattering in order to extract the weak matrix element and to verify the existence of the new negative resonance of $^{204}$Pb nucleus.

## 1. INTRODUCTION

The first theory of weak interaction was proposed by Fermi [1] for the explanation of neutrons and protons decay (or $\beta^\pm$ decay) in atomic nucleus by following reactions:

$$n \rightarrow p + e^- + \nu_e, \quad p \rightarrow n + e^+ + \bar{\nu}_e \tag{1}$$

Another process due to the weak interaction is the muons decay:

$$\mu^+ \to e^+ + \nu_e + \overline{\nu}_\mu, \quad \mu^- \to e^- + \overline{\nu}_e + \nu_\mu \tag{2}$$

The first theory of weak interaction was a parity conserving theory and later was demonstrated that β decay is a process where the parity is not conserving. In 1957 Tanner in his work has published first theoretical work on parity violation in weak interaction between nucleons [2]. In the same year taking into account experimental and theoretical results on weak interaction Feynmann and Gell – Man had emitted the hypothesis of universality of weak interaction [3]. After few years Yu. Abov and collaborators from FLNP – JINR Dubna, in 1964, have measured an asymmetry of emitted gamma quanta in the capture of slow transversal polarized neutrons by $^{113}$Cd nuclei [4]. They have explained the non zero value (of order of $10^{-4}$) of the measured asymmetry by the presence of the weak interaction between nucleons in the compound nucleus $^{114}$Cd, formed by neutron capture. The experimental result of Abov and collaborators gave a serious impulse to theoretical and experimental researches of parity violation effects in nuclear reactions. Parity violation in nuclear reactions is very difficult to measure because this type of interaction acts in the background of nuclear strong interaction which is with order of magnitude higher.

## 2. THEORETICAL BACKGROUNDS

One of the most successful theory of parity violation in nuclear reaction is the formalism proposed by Flambaum and Sushkov [5,6,7]. The main idea of this approach is the following: by interaction of incident slow neutrons with the target nucleus is forming a compound nucleus characterized by spin, parity and others properties and described by resonance states. If a pair of resonances (resonance – resonance approach) has the same spin and opposite parities it is possible to observe PV effects in nuclear reactions with neutrons. In nucleon – nucleon interaction PV effects are of orders of $10^{-7}$ but due to the fact that resonance states are very excited states of nuclei, PV effects are increased with 3-4 orders of magnitude due to the amplification mechanisms (kinematical, dynamical and structural) described in [5,6,7]. The approach of Flambaum and Sushkov is named the formalism of mixing states of the compound nucleus with the same spin and opposite parities. This formalism explained very well the experimental results of Abov and collaborators and others experimental data for different medium and heavy nuclei obtained later in (n,γ), (n,p), (n,α), other reactions and fission.

This formalism proposes the expressions of reaction amplitudes of s and p neutrons in the interaction with target nucleus. Usually neutrons with orbital momentum l=0 are named s-neutrons and by interaction with target nuclei result S-resonances. Neutrons with orbital momentum l=1 are called p-neutrons and by interaction with target nucleus P-resonances are formed. With this introduction the amplitudes of strong interaction with formation of one S and respectively P resonance have the form:

$$f_S^{PC}(E_n) \sim \frac{\sqrt{\Gamma_x^S \Gamma_n^S}}{\left(E - E_S + i\frac{\Gamma_S}{2}\right)} \exp\left[i(\varphi_x^S - \varphi_n^S)\right] \qquad (3)$$

$$f_P^{PC}(E_n) \sim \frac{\sqrt{\Gamma_x^P \Gamma_n^P}}{\left(E - E_P + i\frac{\Gamma_S}{2}\right)} \exp\left[i(\varphi_x^P - \varphi_n^P)\right] \qquad (4)$$

Amplitudes from relations (3) and (4) describe the capture of an s and respectively p neutron with formation of the corresponding S and P resonance and the emission in the exit channel of a particle x (x=n, γ, p, d, t, $^3$He, α,...).

The amplitudes corresponding to the weak interaction not conserving the parity are:

$$f_{S \to P}^{PV}(E_n) \sim W_{SP} \frac{\sqrt{\Gamma_x^P \Gamma_n^S}}{\left(E - E_S + i\frac{\Gamma_S}{2}\right)\left(E - E_P + i\frac{\Gamma_P}{2}\right)} \exp\left[i(\varphi_x^P - \varphi_n^S)\right] \qquad (5)$$

$$f_{P \to S}^{PV}(E_n) \sim W_{SP} \frac{\sqrt{\Gamma_x^S \Gamma_n^P}}{\left(E - E_S + i\frac{\Gamma_S}{2}\right)\left(E - E_P + i\frac{\Gamma_P}{2}\right)} \exp\left[i(\varphi_x^S - \varphi_n^P)\right] \qquad (6)$$

These amplitudes describe the following processes: capture of an s and respectively p neutron with formation of an S and P resonances. In the compound nucleus due to the weak interaction between nucleons the compound nucleus switches between S and P states and in the exit channel an x particle is emitted.

In the relations (3-6) we have $E_S$, $E_P$ = resonance energies of S and P states, $\Gamma_S$, $\Gamma_P$ = total widths in the S and P states of the compound nucleus, $\Gamma_x^S, \Gamma_x^P, \Gamma_n^S, \Gamma_n^P$ = x particle and neutron widths in the S and P states of the compound nucleus, $\varphi_x^S, \varphi_x^P, \varphi_n^S, \varphi_n^P$ = x particle and neutron phases in the S and P states, $W_{SP} = \langle P|H_{weak}|S\rangle$ = weak matrix element, H$_{weak}$ = Hamiltonian of the weak interaction connecting $|S\rangle$ and $|P\rangle$ states of the compound nucleus.

The authors of the present paper have analyzed asymmetry (forward – backward and left - right) and PV (parity non conservation) effects in the $^{35}$Cl(n,p)$^{35}$S reaction with thermal and resonant neutrons. Using existing experimental data of asymmetry and PV effects and their theoretical evaluation in the two-levels approximation, it was extracted the value of the weak matrix element ($W_{SP} = (0.057 \pm 0.0012)$ eV) and were obtained new data on neutron and proton partial reduced widths [8, 9].

A comprehensive understanding and further details of Flambaum – Sushkov formalism can be found in references [5, 6, 7].

## 3. DEFINITIONS OF PV EFFECTS

The present work will deal with PV effects in the scattering of slow neutrons with nuclei. In the case of an incident transversal polarized neutrons beam, in the presence of the weak interaction between nucleons in compound nucleus it is possible to evidence the asymmetry of emitted neutrons, the spin rotation and the spin rotation per unit length. If the incident neutron beam is not polarized or longitudinally polarized then the longitudinal polarization of emitted neutrons can be observed also. The relations of definition for mentioned PV effects are [10, 11]:

- Asymmetry of emitted neutrons, α:

$$\alpha = \frac{\sigma(\Omega,\uparrow) - \sigma(\Omega,\downarrow)}{\sigma(\Omega,\uparrow) + \sigma(\Omega,\downarrow)} \qquad (7)$$

- Spin rotation, Φ:

$$\Phi = \frac{1}{N\sigma_{tot}} \cdot \frac{d\Phi}{dz} = \frac{\text{Re}(f_- - f_+)}{\text{Im}(f_- + f_+)} \qquad (8)$$

- Spin rotation per unit length:

$$\frac{d\Phi}{dz} = N\lambda \,\text{Re}(f_- - f_+) \qquad (9)$$

- Longitudinal polarization, P

$$P = \frac{\sigma_- - \sigma_+}{\sigma_- + \sigma_+} \tag{10}$$

In the above definitions were used the following terms and parameters: $\sigma(\Omega,\uparrow), \sigma(\Omega,\downarrow)$ = differential cross section of scattered neutrons with spins up ($\uparrow$) and down ($\downarrow$); $f_+, f_-$ = scattering amplitude in the forward direction with positive (+) and negative (-) helicity of neutrons; $\sigma_+, \sigma_-$ = total cross section with (+) and (-) neutron helicity; $\sigma_{tot}$ = total cross section, $N$ = number of target nuclei per unit volume, $\lambda$ = neutron wave length.

The PV effects defined in relations (7-10) can be interpreted as correlations betweens vectors from incident and / or emergent channels describing the scattering. According with [10], [11] the asymmetry of scattered (emitted) neutrons ($\alpha$) is the result of the correlation of the type:

$$\alpha \to \left( \vec{\sigma}_i \cdot \vec{n}_f \right) \tag{11}$$

Then for spin rotation we have the following correlation:

$$\Phi, \frac{d\Phi}{dz} \to \left( \vec{\sigma}_i \times \vec{\sigma}_f \right) \cdot \vec{n}_f \tag{12}$$

Finally for polarization (P):

$$P \to \left( \vec{\sigma}_f \cdot \vec{n}_f \right) \tag{13}$$

where $\vec{\sigma}_i, \vec{\sigma}_f, \vec{n}_f$ = unit vectors of the spin of incident and scattered neutrons and direction of scattered neutrons respectively; "$\times, \cdot =$" vector and scalar product.

## 4. RESULTS AND DISCUSSIONS

There are a few experimental data of PV effects on natural Lead by scattering of slow and cold neutrons. The small number of mentioned data is given by the very low value of measured effects and resulting from here the difficulties in the realization of measurements. The only existing experimental data are the spin rotation per unit length and longitudinal polarization.

The natural Lead has four stable isotopes and together with corresponding natural abundance they are: $^{204}$Pb(1.43%), $^{206}$Pb(24.15%), $^{207}$Pb(22.5%) and $^{208}$Pb(52.4%) [12].

The experimental values for spin rotation in the case of cold neutrons scattering ($E_n = 1.76 \cdot 10^{-3}$ eV) are [13]:

$$\frac{d\Phi^{exp}}{dz} = (2.24 \pm 0.33)\cdot 10^{-6}\, rad/cm, \frac{d\Phi^{exp}}{dz} = (3.53 \pm 0.79)\cdot 10^{-6}\, rad/cm \quad (14)$$

For thermal neutron ($E_n = 0.0253$ eV) the experimental value of longitudinal polarization P is [14]:

$$P^{exp} = (-0.7 \pm 0.8)\cdot 10^{-6} \quad (15)$$

The experimental data from (14) and (15) are for natural Lead. In the work [15] it was showed that the main contribution of PV effects in slow neutron scattering on natural Lead is given mainly by $^{204}$Pb isotope therefore it is necessary to make the correction for $^{204}$Pb nucleus with natural abundance 1.34%.

The stable isotope $^{204}$Pb in the region of slow neutrons has two resonances, one negative S and one P resonance with energy, spin and parity respectively [16]:

$$E_S = -2980\, eV, J_S^\pi = \frac{1}{2}^+, E_P = 480\, eV, J_P^\pi = \frac{1}{2}^- \quad (16)$$

In this case, according with resonance-resonance mechanism of PV effects [4, 5, 6], applying the approach suggested in [10, 11], based on the Born and two levels approximations, using the relations of definition (8, 9, 10), the asymmetry of emitted neutron, the spin rotation, spin rotation per unit length and longitudinal polarization have the form:

$$\alpha = 2W_{SP}\sqrt{\Gamma_S^n \Gamma_P^n}\, \frac{\Gamma_S^n(E-E_P) - \Gamma_P^n(E-E_S) + 2kR(E-E_S)(E-E_P)}{(\Gamma_S^n)^2[P] + (\Gamma_P^n)^2[S] + 4(kR)^2[S][P]} \quad (17)$$

$$\Phi = \frac{4W_{SP}\sqrt{\Gamma_S^n \Gamma_P^n}}{[S][P]} \cdot \frac{(E-E_S)(E-E_P) - \dfrac{\Gamma_S^n \Gamma_P^n}{4}}{\dfrac{\Gamma_S \Gamma_S^n}{[S]} + \dfrac{\Gamma_P \Gamma_P^n}{[P]} + 4(kR)^2} \quad (18)$$

$$\frac{d\Phi}{dz} = \frac{N\lambda^2}{\pi} \frac{4W_{SP}\sqrt{\Gamma_S^n \Gamma_P^n}}{[S][P]} \left\{ (E-E_S)(E-E_P) - \frac{\Gamma_S^n \Gamma_P^n}{4} \right\} \quad (19)$$

$$P = -2W_{SP}\sqrt{\Gamma_S^n \Gamma_P^n}\, \frac{(E-E_S)\Gamma_P + (E-E_P)\Gamma_S}{\Gamma_S \Gamma_S^n [P] + \Gamma_P \Gamma_P^n [S] + 4(kR)^2 [S][P]} \quad (20)$$

with $\Gamma_S$, $\Gamma_P$ = total S, P widths, $\Gamma_S^n$, $\Gamma_P^n$ = neutron S, P widths, $E_S$, $E_P$ = S, P resonance energy, $R$ = nucleus radius, $k$ = neutron length number, $[S] = (E-E_S)^2 + 0.25 \cdot \Gamma_S^2$, $[P] = (E-E_P)^2 + 0.25 \cdot \Gamma_P^2$.

Energetic dependences of the expressions (17, 18, 19, 20) are given in Figure 1.

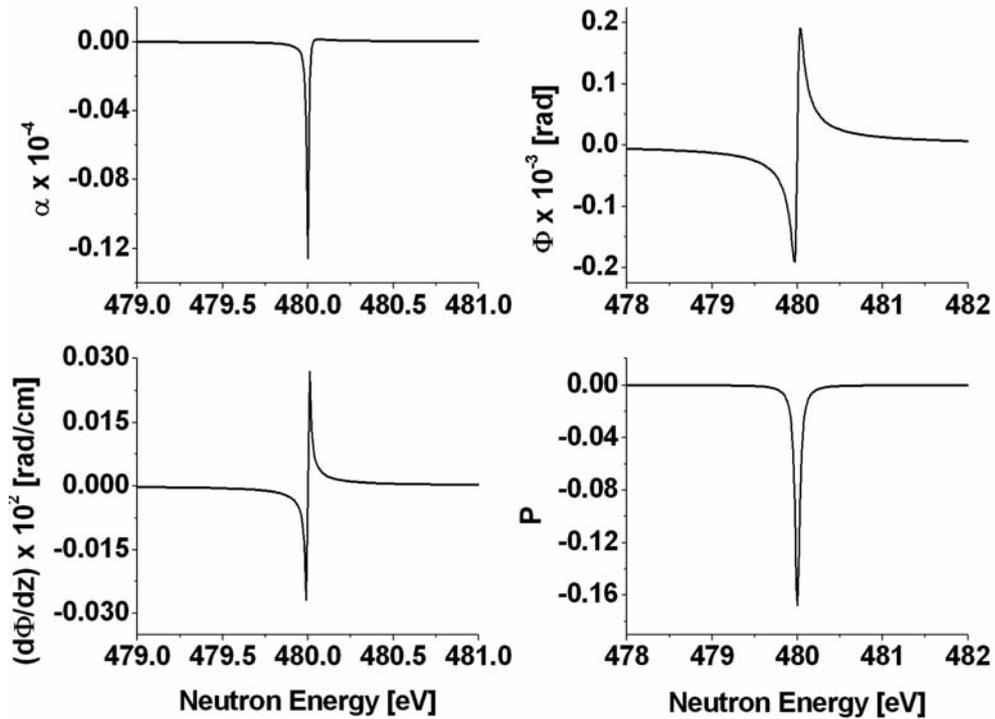

Fig. 1. Energetic dependence of α, Φ, dΦ/dz and P for $^{204}$Pb in the two levels approximation using the resonances $E_S$ = -2980 eV and $E_P$ = 480 eV. The weak matrix element $W_{SP}$ = 2·10$^{-4}$ eV

From Fig.1 results that the studied PV effects have very low value for slow neutrons energies with a consistent modification of order of magnitude near the P resonance as was demonstrated in the fundamental works [5, 6, 7] and [11].

Now the existing experimental data from (14, 15) are compared with theoretical evaluation of the PV effects according with the expressions (17, 18, 19, 20) of $\alpha$, $\Phi$, $d\Phi/dz$ and P for $^{204}$Pb nucleus. For neutrons incident energy, $E_n = 1.76 \cdot 10^{-3}$ eV the spin rotation per unit length is:

$$\frac{d\Phi^{theor}}{dz} = -2.51 \cdot 10^{-8} \, rad/cm \tag{21}$$

The longitudinal polarization P for thermal neutrons ($E_n = 0.0253$ eV) has the value:

$$P^{theor} = -1.5 \cdot 10^{-11} \tag{22}$$

For the asymmetry of emitted neutrons and spin rotation there are not experimental data but we give their theoretical values in order to have in mind the order of magnitude of the PV process.

$$\alpha^{theor} = -1.5 \cdot 10^{-16}, \; \Phi^{theor} = -9.2 \cdot 10^{-8} \, rad \tag{23}$$

In the Fig.1 and in theoretical evaluations (21, 22, 23) the weak matrix element entering in the expressions of PV effects was considered as suggested in [10] for heavy nuclei and has the value, $W_{SP} = 2 \cdot 10^{-4}$ eV.

The longitudinal polarization dependence from Fig.1 shows a high value of order of ten percent in the vicinity of the P resonance. An analog effect was predicted and experimental measured for the first time in the scattering of slow neutrons on $^{139}$La nucleus ($E_P = 0.74$ eV) in LNF JINR Dubna [17, 18].

Comparison between theoretical evaluation and existing experimental data shows a very high discrepancy. It is easy to see that the experimental data are with orders of magnitude higher than theoretical evaluations. In reference [15] the author tried to explain this great difference by the existence of a new negative P resonance, near the neutron binding energy in the compound nucleus, not indicated in the atlas of neutron resonance parameters [16]. After relative simple calculation the author of [15] has proposed a new P resonance with energy $E_P = -16$ eV. Using the expressions (17, 18, 19, 20) of PV effects and the new P resonance energy were obtained the following values:

$$\frac{d\Phi^{theor}}{dz}(E_n = 1.75 \cdot 10^{-3} eV) = 7.54 \cdot 10^{-7} \, rad/cm \tag{24}$$

$$P^{theor}(E_n = 0.0253 eV) = -2.58 \cdot 10^{-8} \tag{25}$$

$$\alpha^{theor} = 4.51 \cdot 10^{-15}, \; \Phi^{theor} = 2.76 \cdot 10^{-6} \, rad \tag{26}$$

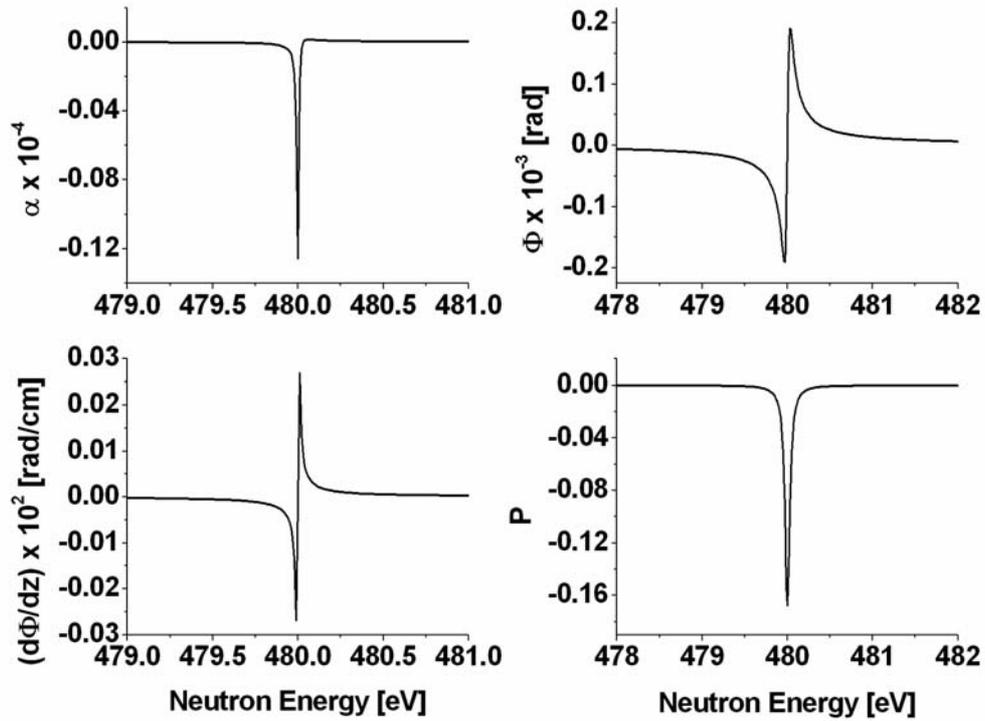

Fig. 2. Energetic dependence of $\alpha$, $\Phi$, $d\Phi/dz$ and $P$ for $^{204}$Pb in the two levels approximation using the resonances $E_S = -2980$ eV and $E_P = -16$ eV. The weak matrix element $W_{SP} = 2 \cdot 10^{-4}$ eV

Introduction of the negative new P resonance leads to the decreasing of the difference between experimental and theoretical data but still the difference remains very big. Qualitatively, between Fig.1 and Fig.2 there is a difference given by the presence of a P resonance with positive energy (Fig. 1) and negative energy (Fig. 2). This difference could be used for future experiments because for a chosen PV effect, a shape of type from Fig. 1 or Fig. 2 can be an answer to the questions of the existence of a new P resonance.

The authors of the present work propose a new method for determination of the existence of a new P resonance. A similar method it was proposed in [9] in the (n,p) reaction on $^{35}$Cl nucleus with slow and resonant neutrons where using only experimental data on forward – backward, left – right and parity non conservation asymmetry effects with their theoretical evaluation, using the two levels approximation in the frame Flambaum – Sushkov formalism [5, 6, 7] the matrix element of the weak non leptonic interaction, $W_{SP}$, was extracted.

In the present situation using the experimental data on spin rotation per unit length and longitudinal polarization (relations (14, 15)) and the corresponding theoretical formulas (19, 20) a system of two equations with two unknown parameters - weak matrix element ($W_{SP}$) and P resonance energy ($E_P$) - is formed.

$$\frac{d\Phi^{teor}}{dz}\left(E_n = 1.76 \cdot 10^{-3} eV, E_P, W_{SP}\right) = \frac{d\Phi^{exp}}{dz} \tag{27}$$

$$P^{teor}\left(E_n = 0.0253 eV, E_P, W_{SP}\right) = P^{exp} \tag{28}$$

The system of equations (27, 28) can be solved numerically and the energy of P resonance and weak matrix element can be extracted. Due to the fact that the experimental value of longitudinal polarization for thermal neutrons is determined with a high error this approach can by considered only qualitatively with a demonstrative character. Still, with the improving of experiments in the future, the data extracted with this new method could be considered quantitatively as well, like in reference [9]. By solving the equations (27, 28) two sets of data were obtained and these are:

$$W_{SP} = 3.7 \cdot 10^{-4} eV, \ E_P = -9 eV \tag{29}$$

$$W_{SP} = 5.6 \cdot 10^{-3} eV, \ E_P = 135 eV \tag{30}$$

The results from (29) and (30) can not answer in a fully way to the questions of the existence of a new P resonance but not exclude this possibility. The second possibility (30) could indicate in fact the P resonance from atlas of neutron resonance parameters with energy $E_P$ = 480 eV "distorted" by experimental errors. In both cases the weak matrix element, $W_{SP}$, is in the limits indicated in [10].

The theoretical evaluations in the present paper were obtained with the help of the computer codes created by the authors. In these codes were implemented the Flambaum – Sushkov approach in the two levels approximation and the new method for the extraction of the weak matrix element and P resonance energy. The computer programs also allow the possibility of exporting the data in ASCII or graphical format.

## 5. CONCLUSIONS

In this work a new approach is proposed for the verification of the presence of a P resonance not indicated in the atlas of neutron resonance parameters [16].

Until now this approach has a qualitative and demonstrative character due to the lack of reliable experimental data and does not exclude the existence of a negative P resonance suggested in [15]. Also in reference [15] it was suggested that the studied PV effects from the present work on natural Lead are given mainly by $^{204}$Pb nucleus. This statement needs to be analyzed once again in future.

The Lead nucleus is a heavy one but due to his structure has no so many neutron resonances. Therefore in our calculations were used only the two levels approximation but the influence of other resonances also could be necessary.

A serious improvement of the method for extraction of the weak matrix element and the energy of new P resonance is the theoretical and experimental evaluation of a similar coefficient to longitudinal polarization of emitted neutrons P, namely, the asymmetry of emitted gamma quanta obtained by the capture of neutrons with opposite helicity. The relation of definition of the asymmetry of emitted gamma quanta is:

$$A_\gamma = \frac{\sigma_{n\gamma}^- - \sigma_{n\gamma}^+}{\sigma_{n\gamma}^- + \sigma_{n\gamma}^+} \qquad (31)$$

where $\sigma_{n\gamma}^+, \sigma_{n\gamma}^-$ = capture cross sections of neutrons with opposite helicity conventionally noted "+" and respectively "-".

In the reference [15] and in the presence work it was tried to explain the big discrepancy between experimental and theoretical data of PV effects in neutron scattering on natural Lead in the frame of resonant – resonant approach of Flambaum – Sushkov and by introducing a new P resonance. The supposition of the existence of new resonances in order to explain PV effects could be considered a little bit forced. The resonant – resonant approach from [5, 6, 7] it is the most known but it is not the only mechanism able to explain the PV phenomena in nuclear reactions. In the theoretical work [10] are proposed other type of PV mechanisms which not require the introduction of new resonances. The analysis of these possibilities must be considered not only for neutron scattering on natural Lead but also for others reactions and nuclei.

This work was supported partially by the Cooperation Program of Plenipotentiary Representative of Romanian Government to JINR Dubna for 2012 year.